%%%%%%%%%%%%%%%%%%%%%%%%%%%%%%%%%%%%%%%%%%%%%%%%%%%%%%%%%%%%%%%%%%%%%%%%%%
%   C-algebras and their applications to reflection groups and cft's     %
%  									 % 
%                        by  J.-B. Zuber                 		 %
%                TEX file, using harvmac.tex macros (included)	         %
%									 %
%                    SPhT  97/077  hep-th/9707034                        %
%%%%%%%%%%%%%%%%%%%%%%%%%%%%%%%%%%%%%%%%%%%%%%%%%%%%%%%%%%%%%%%%%%%%%%%%%%

\overfullrule=0mm
%\input harvmac
%%%%%%%%%%%%%%%%%%  harvmac.tex macros  %%%%%%%%%%%%%%%%%%%%%%%%%%%%%%%%%
%% site dependent options: 
%% \unredoffs and \redoffs define horizontal and vertical offsets 
%% respectively for unreduced and reduced modes. \speclscape defines
%% the \special{} call that sets printer to landscape (sideways) mode.
%% from standard set below, leave uncommented as appropriate or redefine
%
%%% next 400dpi
%\def\unredoffs{} \def\redoffs{\voffset=-.31truein\hoffset=-.48truein}
%\def\speclscape{\special{landscape}}
%
%%% apple lw
%\def\unredoffs{} \def\redoffs{\voffset=-.31truein\hoffset=-.59truein}
%\def\speclscape{\special{ps: landscape}}
%
%%% qms lasergrafix:
%\def\unredoffs{} \def\redoffs{\voffset=-.4truein\hoffset=.125truein}
%\def\speclscape{\special{qms: landscape}}
%
%%% saclay A4 paper:
\def\unredoffs{\hoffset-.14truein\voffset-.2truein} 
\def\redoffs{\voffset=-.45truein\hoffset=-.21truein} 
\def\speclscape{}
%
%---------------------------------------------------------------------%
%
\newbox\leftpage \newdimen\fullhsize \newdimen\hstitle \newdimen\hsbody
\tolerance=1000\hfuzz=2pt
\catcode`\@=11 % This allows us to modify PLAIN macros.
\def\bigans{b }
\def\answ{b }
%\message{ big or little (b/l)? }\read-1 to\answ
%
\ifx\answ\bigans\message{(This will come out unreduced.}
\magnification=1200\unredoffs\baselineskip=16pt plus 2pt minus 1pt
\hsbody=\hsize \hstitle=\hsize %take default values for unreduced format
\else\message{(This will be reduced.} \let\l@r=L
\magnification=1000\baselineskip=16pt plus 2pt minus 1pt \vsize=7truein
\redoffs \hstitle=8truein\hsbody=4.75truein\fullhsize=10truein\hsize=\hsbody
\output={\ifnum\pageno=0 %%% This is the HUTP version
  \shipout\vbox{\speclscape{\hsize\fullhsize\makeheadline}
    \hbox to \fullhsize{\hfill\pagebody\hfill}}\advancepageno
  \else
  \almostshipout{\leftline{\vbox{\pagebody\makefootline}}}\advancepageno 
  \fi}
\def\almostshipout#1{\if L\l@r \count1=1 \message{[\the\count0.\the\count1]}
      \global\setbox\leftpage=#1 \global\let\l@r=R
 \else \count1=2
  \shipout\vbox{\speclscape{\hsize\fullhsize\makeheadline}
      \hbox to\fullhsize{\box\leftpage\hfil#1}}  \global\let\l@r=L\fi}
\fi
%---------------------------------------------------------------------
%
\newcount\yearltd\yearltd=\year\advance\yearltd by -1900

\def\Title#1#2{\nopagenumbers\abstractfont\hsize=\hstitle\rightline{#1}%
\vskip 1in\centerline{\titlefont #2}\abstractfont\vskip .5in\pageno=0}
\def\Date#1{\vfill\leftline{#1}\tenpoint\supereject\global\hsize=\hsbody%
\footline={\hss\tenrm\folio\hss}}% 	restores pagenumbers
%
%       use following instead of \Date on the preliminary draft, 
%       puts date/time on each page in big mode, writes labels in margins

\def\draftmode{\message{ DRAFTMODE }\def\draftdate{{\rm preliminary draft:
\number\month/\number\day/\number\yearltd\ \ \hourmin}}%
\headline={\hfil\draftdate}\writelabels\baselineskip=20pt plus 2pt minus 2pt
 {\count255=\time\divide\count255 by 60 \xdef\hourmin{\number\count255}
  \multiply\count255 by-60\advance\count255 by\time
  \xdef\hourmin{\hourmin:\ifnum\count255<10 0\fi\the\count255}}}
%       use \nolabels to get rid of eqn, ref, and fig labels in draft mode
\def\nolabels{\def\wrlabeL##1{}\def\eqlabeL##1{}\def\reflabeL##1{}}
\def\writelabels{\def\wrlabeL##1{\leavevmode\vadjust{\rlap{\smash%
{\line{{\escapechar=` \hfill\rlap{\sevenrm\hskip.03in\string##1}}}}}}}%
\def\eqlabeL##1{{\escapechar-1\rlap{\sevenrm\hskip.05in\string##1}}}%
\def\reflabeL##1{\noexpand\llap{\noexpand\sevenrm\string\string\string##1}}}
\nolabels
%
% tagged sec numbers
\global\newcount\secno \global\secno=0
\global\newcount\meqno \global\meqno=1
\def\newsec#1{\global\advance\secno by1\message{(\the\secno. #1)}
%\ifx\answ\bigans \vfill\eject \else \bigbreak\bigskip \fi  %if desired
\global\subsecno=0\eqnres@t\noindent{\bf\the\secno. #1}
\writetoca{{\secsym} {#1}}\par\nobreak\medskip\nobreak}
\def\eqnres@t{\xdef\secsym{\the\secno.}\global\meqno=1\bigbreak\bigskip}
\def\sequentialequations{\def\eqnres@t{\bigbreak}}\xdef\secsym{}
\global\newcount\subsecno \global\subsecno=0
\def\subsec#1{\global\advance\subsecno by1\message{(\secsym\the\subsecno. #1)}
\ifnum\lastpenalty>9000\else\bigbreak\fi
\noindent{\it\secsym\the\subsecno. #1}\writetoca{\string\quad 
{\secsym\the\subsecno.} {#1}}\par\nobreak\medskip\nobreak}
\def\appendix#1#2{\global\meqno=1\global\subsecno=0\xdef\secsym{\hbox{#1.}}
\bigbreak\bigskip\noindent{\bf Appendix #1. #2}\message{(#1. #2)}
\writetoca{Appendix {#1.} {#2}}\par\nobreak\medskip\nobreak}
%
%       \eqn\label{a+b=c}	gives displayed equation, numbered
%				consecutively within sections.
%     \eqnn and \eqna define labels in advance (of eqalign?)
%
\def\eqnn#1{\xdef #1{(\secsym\the\meqno)}\writedef{#1\leftbracket#1}%
\global\advance\meqno by1\wrlabeL#1}
\def\eqna#1{\xdef #1##1{\hbox{$(\secsym\the\meqno##1)$}}
\writedef{#1\numbersign1\leftbracket#1{\numbersign1}}%
\global\advance\meqno by1\wrlabeL{#1$\{\}$}}
\def\eqn#1#2{\xdef #1{(\secsym\the\meqno)}\writedef{#1\leftbracket#1}%
\global\advance\meqno by1$$#2\eqno#1\eqlabeL#1$$}
%
%			 footnotes
\newskip\footskip\footskip14pt plus 1pt minus 1pt %sets footnote baselineskip
\def\footnotefont{\ninepoint}\def\f@t#1{\footnotefont #1\@foot}
\def\f@@t{\baselineskip\footskip\bgroup\footnotefont\aftergroup\@foot\let\next}
\setbox\strutbox=\hbox{\vrule height9.5pt depth4.5pt width0pt}
\global\newcount\ftno \global\ftno=0
\def\foot{\global\advance\ftno by1\footnote{$^{\the\ftno}$}}
%
%say \footend to put footnotes at end
%will cause problems if \ref used inside \foot, instead use \nref before
\newwrite\ftfile   
\def\footend{\def\foot{\global\advance\ftno by1\chardef\wfile=\ftfile
$^{\the\ftno}$\ifnum\ftno=1\immediate\openout\ftfile=foots.tmp\fi%
\immediate\write\ftfile{\noexpand\smallskip%
\noexpand\item{f\the\ftno:\ }\pctsign}\findarg}%
\def\footatend{\vfill\eject\immediate\closeout\ftfile{\parindent=20pt
\centerline{\bf Footnotes}\nobreak\bigskip\input foots.tmp }}}
\def\footatend{}
%
%     \ref\label{text}
% generates a number, assigns it to \label, generates an entry.
% To list the refs on a separate page,  \listrefs
%
\global\newcount\refno \global\refno=1
\newwrite\rfile
\def\ref{[\the\refno]\nref}
\def\nref#1{\xdef#1{[\the\refno]}\writedef{#1\leftbracket#1}%
\ifnum\refno=1\immediate\openout\rfile=refs.tmp\fi
\global\advance\refno by1\chardef\wfile=\rfile\immediate
\write\rfile{\noexpand\item{#1\ }\reflabeL{#1\hskip.31in}\pctsign}\findarg}
%	horrible hack to sidestep tex \write limitation
\def\findarg#1#{\begingroup\obeylines\newlinechar=`\^^M\pass@rg}
{\obeylines\gdef\pass@rg#1{\writ@line\relax #1^^M\hbox{}^^M}%
\gdef\writ@line#1^^M{\expandafter\toks0\expandafter{\striprel@x #1}%
\edef\next{\the\toks0}\ifx\next\em@rk\let\next=\endgroup\else\ifx\next\empty%
\else\immediate\write\wfile{\the\toks0}\fi\let\next=\writ@line\fi\next\relax}}
\def\striprel@x#1{} \def\em@rk{\hbox{}} 
\def\lref{\begingroup\obeylines\lr@f}
\def\lr@f#1#2{\gdef#1{\ref#1{#2}}\endgroup\unskip}
\def\semi{;\hfil\break}
\def\addref#1{\immediate\write\rfile{\noexpand\item{}#1}} %now unnecessary
\def\startrefs#1{\immediate\openout\rfile=refs.tmp\refno=#1}
\def\xref{\expandafter\xr@f}\def\xr@f[#1]{#1}
\def\refs#1{\count255=1[\r@fs #1{\hbox{}}]}
\def\r@fs#1{\ifx\und@fined#1\message{reflabel \string#1 is undefined.}%
\nref#1{need to supply reference \string#1.}\fi%
\vphantom{\hphantom{#1}}\edef\next{#1}\ifx\next\em@rk\def\next{}%
\else\ifx\next#1\ifodd\count255\relax\xref#1\count255=0\fi%
\else#1\count255=1\fi\let\next=\r@fs\fi\next}
%

%
% this is ugly, but moore insists
\newwrite\ffile\global\newcount\figno \global\figno=1
\def\fig{fig.~\the\figno\nfig}
\def\nfig#1{\xdef#1{fig.~\the\figno}%
\writedef{#1\leftbracket fig.\noexpand~\the\figno}%
\ifnum\figno=1\immediate\openout\ffile=figs.tmp\fi\chardef\wfile=\ffile%
\immediate\write\ffile{\noexpand\medskip\noexpand\item{Fig.\ \the\figno. }
\reflabeL{#1\hskip.55in}\pctsign}\global\advance\figno by1\findarg}
\def\xfig{\expandafter\xf@g}\def\xf@g fig.\penalty\@M\ {}
\def\figs#1{figs.~\f@gs #1{\hbox{}}}
\def\f@gs#1{\edef\next{#1}\ifx\next\em@rk\def\next{}\else
\ifx\next#1\xfig #1\else#1\fi\let\next=\f@gs\fi\next}
\newwrite\lfile
{\escapechar-1\xdef\pctsign{\string\%}\xdef\leftbracket{\string\{}
\xdef\rightbracket{\string\}}\xdef\numbersign{\string\#}}

\def\writestop{\def\writestoppt{\immediate\write\lfile{\string\pageno%
\the\pageno\string\startrefs\leftbracket\the\refno\rightbracket%
\string\def\string\secsym\leftbracket\secsym\rightbracket%
\string\secno\the\secno\string\meqno\the\meqno}\immediate\closeout\lfile}}
\def\writestoppt{}\def\writedef#1{}
\def\seclab#1{\xdef #1{\the\secno}\writedef{#1\leftbracket#1}\wrlabeL{#1=#1}}
\def\subseclab#1{\xdef #1{\secsym\the\subsecno}%
\writedef{#1\leftbracket#1}\wrlabeL{#1=#1}}
\newwrite\tfile \def\writetoca#1{}
\def\leaderfill{\leaders\hbox to 1em{\hss.\hss}\hfill}
%	use this to write file with table of contents
\def\writetoc{\immediate\openout\tfile=toc.tmp 
   \def\writetoca##1{{\edef\next{\write\tfile{\noindent ##1 
   \string\leaderfill {\noexpand\number\pageno} \par}}\next}}}
%       and this lists table of contents on second pass

%
\catcode`\@=12 % at signs are no longer letters
%
%	Unpleasantness in calling in abstract and title fonts
\edef\tfontsize{\ifx\answ\bigans scaled\magstep3\else scaled\magstep4\fi}
\font\titlerm=cmr10 \tfontsize \font\titlerms=cmr7 \tfontsize
\font\titlermss=cmr5 \tfontsize \font\titlei=cmmi10 \tfontsize
\font\titleis=cmmi7 \tfontsize \font\titleiss=cmmi5 \tfontsize
\font\titlesy=cmsy10 \tfontsize \font\titlesys=cmsy7 \tfontsize
\font\titlesyss=cmsy5 \tfontsize \font\titleit=cmti10 \tfontsize
\skewchar\titlei='177 \skewchar\titleis='177 \skewchar\titleiss='177
\skewchar\titlesy='60 \skewchar\titlesys='60 \skewchar\titlesyss='60
\def\titlefont{\def\rm{\fam0\titlerm}% switch to title font
\textfont0=\titlerm \scriptfont0=\titlerms \scriptscriptfont0=\titlermss
\textfont1=\titlei \scriptfont1=\titleis \scriptscriptfont1=\titleiss
\textfont2=\titlesy \scriptfont2=\titlesys \scriptscriptfont2=\titlesyss
\textfont\itfam=\titleit \def\it{\fam\itfam\titleit}\rm}
 \ifx\answ\bigans\else scaled\magstep1\fi
\ifx\answ\bigans\def\abstractfont{\tenpoint}\else
\font\abssl=cmsl10 scaled \magstep1
\font\absrm=cmr10 scaled\magstep1 \font\absrms=cmr7 scaled\magstep1
\font\absrmss=cmr5 scaled\magstep1 \font\absi=cmmi10 scaled\magstep1
\font\absis=cmmi7 scaled\magstep1 \font\absiss=cmmi5 scaled\magstep1
\font\abssy=cmsy10 scaled\magstep1 \font\abssys=cmsy7 scaled\magstep1
\font\abssyss=cmsy5 scaled\magstep1 \font\absbf=cmbx10 scaled\magstep1
\skewchar\absi='177 \skewchar\absis='177 \skewchar\absiss='177
\skewchar\abssy='60 \skewchar\abssys='60 \skewchar\abssyss='60
\def\abstractfont{\def\rm{\fam0\absrm}% switch to abstract font
\textfont0=\absrm \scriptfont0=\absrms \scriptscriptfont0=\absrmss
\textfont1=\absi \scriptfont1=\absis \scriptscriptfont1=\absiss
\textfont2=\abssy \scriptfont2=\abssys \scriptscriptfont2=\abssyss
\textfont\itfam=\bigit \def\it{\fam\itfam\bigit}\def\footnotefont{\tenpoint}%
\textfont\slfam=\abssl \def\sl{\fam\slfam\abssl}%
\textfont\bffam=\absbf \def\bf{\fam\bffam\absbf}\rm}\fi
\def\tenpoint{\def\rm{\fam0\tenrm}% switch back to 10-point type
\textfont0=\tenrm \scriptfont0=\sevenrm \scriptscriptfont0=\fiverm
\textfont1=\teni  \scriptfont1=\seveni  \scriptscriptfont1=\fivei
\textfont2=\tensy \scriptfont2=\sevensy \scriptscriptfont2=\fivesy
\textfont\itfam=\tenit \def\it{\fam\itfam\tenit}\def\footnotefont{\ninepoint}%
\textfont\bffam=\tenbf \def\bf{\fam\bffam\tenbf}\def\sl{\fam\slfam\tensl}\rm}
\font\ninerm=cmr9 \font\sixrm=cmr6 \font\ninei=cmmi9 \font\sixi=cmmi6 
\font\ninesy=cmsy9 \font\sixsy=cmsy6 \font\ninebf=cmbx9 
\font\nineit=cmti9 \font\ninesl=cmsl9 \skewchar\ninei='177
\skewchar\sixi='177 \skewchar\ninesy='60 \skewchar\sixsy='60 
\def\ninepoint{\def\rm{\fam0\ninerm}% switch to footnote font
\textfont0=\ninerm \scriptfont0=\sixrm \scriptscriptfont0=\fiverm
\textfont1=\ninei \scriptfont1=\sixi \scriptscriptfont1=\fivei
\textfont2=\ninesy \scriptfont2=\sixsy \scriptscriptfont2=\fivesy
\textfont\itfam=\ninei \def\it{\fam\itfam\nineit}\def\sl{\fam\slfam\ninesl}%
\textfont\bffam=\ninebf \def\bf{\fam\bffam\ninebf}\rm} 
%
%---------------------------------------------------------------------
%

\hyphenation{anom-aly anom-alies coun-ter-term coun-ter-terms}
\def\inv{^{\raise.15ex\hbox{${\scriptscriptstyle -}$}\kern-.05em 1}}

\def\Dsl{\,\raise.15ex\hbox{/}\mkern-13.5mu D} %this one can be subscripted
\def\dsl{\raise.15ex\hbox{/}\kern-.57em\partial}

\font\bigit=cmti10 scaled \magstep1
 %pound sterling
\def\lspace{\ifx\answ\bigans{}\else\qquad\fi}
\def\lbspace{\ifx\answ\bigans{}\else\hskip-.2in\fi} % $$\lbspace...$$
\def\boxeqn#1{\vcenter{\vbox{\hrule\hbox{\vrule\kern3pt\vbox{\kern3pt
	\hbox{${\displaystyle #1}$}\kern3pt}\kern3pt\vrule}\hrule}}}
\def\mbox#1#2{\vcenter{\hrule \hbox{\vrule height#2in
		\kern#1in \vrule} \hrule}}  %e.g. \mbox{.1}{.1}
%	matters of taste
%\def\tilde{\widetilde} \def\bar{\overline} \def\hat{\widehat}
%
% some sample definitions
  %     curly letters

\def\darr#1{\raise1.5ex\hbox{$\leftrightarrow$}\mkern-16.5mu #1}
 %pound sterling

 %puts a small half in a displayed eqn
\def\roughly#1{\raise.3ex\hbox{$#1$\kern-.75em\lower1ex\hbox{$\sim$}}}

%%%%%%%%%%%%%%%%%%%%  end of harvmac.tex  %%%%%%%%%%%%%%%%%%%%%%%%%%

%

%Macros 
%%%%%%%%%%%%%%%%%%%%%%%%%%%%%%%%%%%%%%%%%%%%%%%%%%%%%%%%%%%%%%%
%%%%%%%%%%%%%%%%%%%DEFINITIONS%%%%%%%%%%%%%%%%%%%%%%%%%%%%%%%%%
%
\def\frac#1#2{\scriptstyle{#1 \over #2}}		
\def\inv#1{\scriptstyle{1 \over #1}}

%
%%%%%%%%%%%%%%%%%%%CALLIGRAPHIC LETTERS%%%%%%%%%%%%%%%%%%%%%%%%%
%
	\def\hT{{\cal B}}

	\def\CN{{\cal N}}

\def\({ \left( }\def\[{ \left[ }
\def\){ \right) }\def\]{ \right] }
%%%%%%%%%%%%%%%%%%%%MATH CHARACTERS%%%%%%%%%%%%%%%%%%%%%%%%%%%%
%

%\font\numbers=cmu10 scaled\magstep1

\def\IR{\relax{\rm I\kern-.18em R}}
\font\cmss=cmss10 \font\cmsss=cmss10 at 7pt
\def\IZ{\relax\ifmmode\mathchoice
{\hbox{\cmss Z\kern-.4em Z}}{\hbox{\cmss Z\kern-.4em Z}}
{\lower.9pt\hbox{\cmsss Z\kern-.4em Z}}
{\lower1.2pt\hbox{\cmsss Z\kern-.4em Z}}\else{\cmss Z\kern-.4em Z}\fi}
\def\inbar{\,\vrule height1.5ex width.4pt depth0pt}
\def\IB{\relax{\rm I\kern-.18em B}}
\def\IC{\relax\hbox{$\inbar\kern-.3em{\rm C}$}}
\def\ID{\relax{\rm I\kern-.18em D}}
\def\IE{\relax{\rm I\kern-.18em E}}
\def\IF{\relax{\rm I\kern-.18em F}}
\def\IG{\relax\hbox{$\inbar\kern-.3em{\rm G}$}}
\def\IH{\relax{\rm I\kern-.18em H}}
\def\II{\relax{\rm I\kern-.18em I}}
\def\IK{\relax{\rm I\kern-.18em K}}
\def\IL{\relax{\rm I\kern-.18em L}}
\def\IM{\relax{\rm I\kern-.18em M}}
\def\IN{\relax{\rm I\kern-.18em N}}
\def\IO{\relax\hbox{$\inbar\kern-.3em{\rm O}$}}
\def\IP{\relax{\rm I\kern-.18em P}}
\def\IQ{\relax\hbox{$\inbar\kern-.3em{\rm Q}$}}
\def\IGa{\relax\hbox{${\rm I}\kern-.18em\Gamma$}}
\def\IPi{\relax\hbox{${\rm I}\kern-.18em\Pi$}}
\def\ITh{\relax\hbox{$\inbar\kern-.3em\Theta$}}
\def\IOm{\relax\hbox{$\inbar\kern-3.00pt\Omega$}}

%%%%%%%%%%%%%%%%%%%%%%%%%%%%%%%%%%%%%%%%%%%%%%%%%%%%%%%%%%%%%%%%%

%Mes Macros

%\def\Z{_N Z}

%%%%%%%%%%%%%%%%%%%%Greek letters%%%%%%%%%%%%%%%%%%%%%%%%%%%%%%%%%%
\def\Ga{\alpha}\def\Gb{\beta}\def\Gc{\gamma}\def\GC{\Gamma}
\def\Gd{\delta}

\def\Gl{\lambda}
\def\Gm{\mu}\def\Gn{\nu}
\def\Gr{\rho}

%%%%%%%%%%%%%%%% Boldface letters %%%%%%%%%%%%%%%%%%%%%%%%%%%%%%%%
\def\bN{{\bf N}}\def\bD{{\bf D}}\def\bS{{\bf S}}\def\bp{{\bf p}}
%%%%%%%%%%%%%%%% hat letters %%%%%%%%%%%%%%%%%%%%%%%%%%%%%%%%%%%%%%
\def\hk{\hat k }
\def\hT{\hat T }

 \def\Che{Chebishev\ } % pas Che Guevara !!
 
%

%nouvelles macros de ce papier
\def\bt{{\bf t}}
\def\deg{{\rm deg}}
\def\chih{{\widehat \chi}}
\def\nind{{\par\noindent}}
\def\slh{\widehat{sl}}

\catcode`\@=11
\def\Eqalign#1{\null\,\vcenter{\openup\jot\m@th\ialign{
\strut\hfil$\displaystyle{##}$&$\displaystyle{{}##}$\hfil
&&\quad\strut\hfil$\displaystyle{##}$&$\displaystyle{{}##}$
\hfil\crcr#1\crcr}}\,}    \catcode`\@=12

\message{You are asumed to  have the AMS fonts otherwise (un)comment 
the lines below } 

%% If you don't have the AMS fonts comment  the 5 lines below
\input amssym.def
\input amssym.tex
\def\IZ{\Bbb Z}\def\IR{\Bbb R}\def\IC{\Bbb C}\def\IN{\Bbb N}
\def\gg{\goth g} 
\def\gA{\goth A}

%% If you don't have the AMS fonts UNcomment the 2 lines below 
%\def\gg{g}
%\def\gA{\CA}

\def\ggh{\hat\gg}

%%%%%%%%%%%%%%%%%%%%%%%%%%%%%%%%%%%%%%%%%%%%%%%%%%%%%%%%%%%%%%%

\input epsf.tex
%%%%%%%%%%%%%%%%%%%%%% macros for figures %%%%%%%%%%%%%%%%%%%%%%%%%%%
\newcount\figno
\figno=0
\def\fig#1#2#3{
\par\begingroup\parindent=0pt\leftskip=1cm\rightskip=1cm\parindent=0pt
\baselineskip=11pt
\global\advance\figno by 1
\midinsert
\epsfxsize=#3    %epsfysize=vertical size !!
\centerline{\epsfbox{#2}}
\vskip 12pt
{\bf Fig. \the\figno:} #1\par
\endinsert\endgroup\par
}
\def\figlabel#1{\xdef#1{\the\figno}}
\def\encadremath#1{\vbox{\hrule\hbox{\vrule\kern8pt\vbox{\kern8pt
\hbox{$\displaystyle #1$}\kern8pt}
\kern8pt\vrule}\hrule}}

%%%%%%%%%%%%%%%%%%%%%%%%%%%%%%%%%%%%%%%%%%%%%%%%%%%%%%%%%%%%%%%

\Title{\vbox{\hbox{SPhT 97/077}\hbox{{\tt hep-th/9707034   }}}}
%\Title{SPhT 97/077;  hep-th/yymmxxx}
%\Title{}
{{\vbox {
%\centerline{}
\centerline{  C-algebras and their applications } 
\bigskip
\centerline{to reflection groups and conformal field theories 
} }}}

\bigskip

\centerline{J.-B. Zuber}\bigskip

\centerline{\it CEA SACLAY  Service de Physique Th\'eorique de Saclay,}
\centerline{ \it F-91191 Gif sur Yvette Cedex, France}
\vskip .2in

\noindent 
The aim of this lecture is to present the concept of C-algebra
and to illustrate its applications in two contexts: the study 
of reflection groups and their folding on the one hand, 
the structure of rational conformal field theories on the 
other. For simplicity the discussion is restricted to 
finite Coxeter groups and conformal theories with a $\widehat{sl}(2)$
current algebra, but it may be extended to a larger class 
of groups and theories associated with $\widehat{sl}(N)$.

%\Date{}
\Date{4/97\qquad\qquad to appear in  the proceedings of the RIMS Symposium, Kyoto, 16-19 December 1996.}
%\draftmode
%

%%%%

% References
\lref\No{M. Noumi, {\it Tokyo J. Math.} {\bf 7} (1984) 1-60.}
\lref\DFLZ{
P. Di Francesco, F. Lesage and J.-B. Zuber, 
        {\it Nucl. Phys.} {\bf B408} (1993) [FS] 600-634:  hep-th/9306018.}

\lref\CIZ{A. Cappelli, C. Itzykson, and J.-B. Zuber,
{\it Nucl. Phys.} {\bf B280} (1987) [FS18]
445-465; {\it Comm. Math. Phys.} {\bf 113} (1987) 1-26. }
\lref\Ka{A. Kato, {\it Mod. Phys. Lett.} {\bf A2} (1987) 585-600.}

\lref\OEK{A. Ocneanu in 
{\it Operator Algebras and Applications}, vol.2, 119-172, London Math. Soc. 
Lecture Notes Series, Vol. 136, Cambridge Univ. Press, London 1988; 
{\it Quantum symmetry, differential geometry of finite graphs and 
classification of subfactors}, Univ. of Tokyo Seminar Notes 45, notes taken by
Y. Kawahigashi
\semi  Y. Kawahigashi, %{\it On flatness of Oneanu's connections on the 
%Dynkin diagram and classification of subfactors}, preprint 1990
{\it J. Funct. Anal.} {\bf 127}  63-107\semi
D. Evans and Y. Kawahigashi, 
%{\it The $E_7$ commuting squares produce $D_{10}$ as principal graph}, 
% preprint 1992
{\it Pub. RIMS, Kyoto Univ.} {\bf 30} 151-166; 
 {\it Quantum Symmetries in Operator Algebras}, Oxford Univ. Press 
to appear.}

\lref\MB{M. Bauer, unpublished.}

\lref\EV{E. Verlinde, {\it Nucl. Phys.} {\bf B300} [FS22] (1988) 360-376. }
\lref\DGe{D. Gepner, {\it Comm. Math. Phys.} {\bf 141} (1991) 381-411.}
\lref\LW{W. Lerche and N.P. Warner, %proceedings of Stony-Brook
in {\it Strings \& Symmetries, 1991}, N. Berkovits, H. Itoyama et al. eds, 
World Scientific 1992.}

\lref\Pas{V. Pasquier, J.Phys. {\bf A20} 5707-5717 (1987); 
Th\`ese d'Etat, Orsay, 1988. }

\lref\PZun{V.B. Petkova and J.-B. Zuber, 
{\it Nucl. Phys.} {\bf B438} (1995) 347-372:  hep-th/9410209. }

\lref\BI{E. Bannai, T. Ito, {\it Algebraic Combinatorics I: Association
Schemes}, Benjamin/Cummings (1984).}

\lref\PZde{V. Petkova and J.-B. Zuber, 
%{\it From CFT to Graphs}, preprint ASI-TPA/14/95, SPhT~95/118, 
{\it Nucl. Phys. B} {\bf B463} (1996) 161-193: hep-th 9510175; 
{\it Conformal Field Theory and Graphs}, hep-th/9701103, 
to appear in the proceedings of the 21st International 
Colloquium on Group Theoretical Methods in Physics, Goslar, Germany,
 July 1996. }

\lref\CMP{J.-B. Zuber, {\it Comm. Math. Phys.} {\bf 179} (1996) 265-294.}

\lref\Tani{J.-B. Zuber, {\it Generalized Dynkin diagrams and root
systems and their folding}, to appear in 
proceedings of the Taniguchi Symposium
{Topological Field Theory, Primitive Forms and Related Topics}, 
 Kyoto Dec 1996, M. Kashiwara, A. Matsuo,  K. Saito and I. Satake eds, 
Birkha\"user.}

\lref\Yano{J. Sekiguchi and T. Yano, 
{\it Sci. Rep. Saitama Univ.} {\bf IX} (1980) 33-44; 
T. Yano, %{\it Sci. Rep. Saitama Univ.} 
{\it ibid.} 61-70.}
\lref\fold{O.P. Shcherbak, {\it Russ. Math. Surveys} 
{\bf 43:3}  (1988) 149-194 \semi
R.V. Moody and J. Patera, {\it J.Phys.A} {\bf 26} (1993) 2829-2853.} 

\lref\Dub{B. Dubrovin,
{\it Nucl. Phys.} {\bf B 379} (1992) 627-689:
%{\it Differential Geometry of the space of orbits of a reflection group}
{ hep-th/9303152};
%{\it Geometry of 2D Topological Field Theories}, 
Springer Lect. Notes in Math. {\bf 1620} (1996) 120-348: 
{hep-th/9407018};
 B. Dubrovin and Y. Zhang,  
{\it Extended affine Weyl groups and Frobenius manifolds}, hep-th/9611200.}

\lref\Zdub{J.-B. Zuber, {\it Mod. Phys. Lett. A} {\bf 8} (1994) 749-760: 
hep-th/9312209. }

\lref\DFZ{P. Di Francesco and J.-B. Zuber, 
in {\it Recent Developments in Conformal Field Theories}, Trieste
Conference, 1989, S. Randjbar-Daemi, E. Sezgin and J.-B. Zuber eds., 
World Scientific 1990  \semi %.}
%\lref\DF{
 P. Di Francesco, {\it Int.J.Mod.Phys.} {\bf A7} (1992) 407-500.}

\lref\DFZun{P. Di Francesco and J.-B. Zuber, 
{\it Nucl. Phys.} {\bf B338} (1990) 602-646.} 

\lref\WDVV{
E. Witten, Nucl. Phys. {\bf B 340} (1990) 281-322\semi 
R. Dijkgraaf, E. Verlinde and H. Verlinde, Nucl. Phys. 
{\bf B352} (1991) 59-86; 
in {\it String Theory and Quantum Gravity},
proceedings of the 11990 Trieste Spring School, M. Green et al. {\it eds.},
World Sc. 1991.}

\lref\MSDV{R. Dijkgraaf and E. Verlinde, {\it Nucl. Phys.} (Proc. Suppl.)
{\bf 5B}  (1988) 87-97 %.}
\semi G. Moore and N. Seiberg,
%\lref\MS{G. Moore and N. Seiberg,  
{\it Nucl. Phys.} {\bf B313}
(1989) 16-40; {\it Comm. Math. Phys.} {\bf 123} (1989) 177-254.}

%%%%%%%%%%%%%%%%%%%%%%%%%%%%%%%%%%%%%%%%%%%%%%%%%%%%%%%%%%%%

\newsec{Introduction}
\noindent
The purpose of this talk is to present the notion of C-algebra, 
a concept that appears particularly suited in the discussion 
of various topics of current interest in mathematics and 
mathematical physics:  rational conformal
field theories (rcft), topological field theories, singularity 
theory and related problems. The concept was originally developed
in relation with finite groups and the algebras of their
characters and classes (whence the ``C"): this exposes clearly 
one of the key features of these algebras, namely the pattern 
of two dual algebras. More generally, (the precise definition will
be given in sect. 3), C-algebras are associative, commutative algebras
with a finite number of generators. They come in dual pairs, 
endowed with  different multiplication laws, 
one algebra being generated by the idempotents of the other.
 We shall illustrate and apply this concept
in two different contexts:  the association between  rational 
conformal field theories and graphs on the one hand;  the 
folding of root systems and Dynkin diagrams on the other. In both 
cases, generalized Dynkin diagrams are the central objects, and 
pairs of algebras that are naturally associated with these graphs
are C-algebras. The study of the C-subalgebras (to be also
defined below) then enables one to understand the relationship
between rcft and graph --how to construct one object from the other--
and to understand the folding of root systems, Dynkin diagrams and
reflection groups. 

Because certain positivity properties play an important role in the 
discussion of C-algebras, we start with a presentation of such 
 properties that are empirically observed in different contexts
 but do not seem to have been given enough attention. 

For the sake of brevity, all the discussion will be restricted 
to the simplest --and best understood-- case: rcft associated
with $sl(2)$,  ``minimal" topological field theories, 
simple singularities, ordinary Coxeter-Dynkin diagrams, etc.
There is ample evidence, however, --and a few proofs--, 
that the present considerations extend to a much larger context.

\bigskip
%%%%%%%%%%%%%%%%%%%%%%%%%%%%%%%%%%%%%%%%%%%%%%%%%%%%%%%%%%%%%%%%

\newsec{Three empirical facts}
\noindent Consider the prepotential $F(\bt)$ of one of the $ADE$ 
singularities. 
Here $\bt=(t^1,\cdots, t^n)$, where $n$ is the rank of the associated 
 $ADE$ algebra (the Milnor number of the singularity); the $t^j$ are the 
flat coordinates in the versal deformation of the singularity. $F(\bt)$
satisfies the Witten-Dijkgraaf-Verlinde-Verlinde (WDVV) equations \WDVV, 
which express the associativity of the algebra with structure constants
$C_{ij}^{\ \ k}(\bt)$, where $C_{ijk}(\bt)
 ={\partial^3F\over \partial t^i \partial t^j 
\partial t^k}$, and indices are raised and lowered with the $t$-independent
tensor $\eta_{kl}=C_{1kl}$, and its inverse $\eta^{jk}$. The prepotential  
$F$ is a quasihomogeneous polynomial of degree $2(h+1)$ if we assign 
% case of A_n: d(F)=2(n+2); d(t_i)=n+2-i, i=1, ..., n
to the variables $t^i$ the degrees $\deg(t^i)=(h+1-\Gl_i)$, where $\Gl_i$ 
is the $i$-th Coxeter exponent of the $ADE$ algebra: these 
exponents are supposed to be labelled  in increasing order: 
$\Gl_1=1 < \Gl_2 \le \cdots \le\Gl_{n-1}< \Gl_n=h-1$,  
$h$ the Coxeter number. 
It is convenient 
to change notations, labelling the $t$'s with the value of $\Gl_i$, 
hence replacing $t^i$ with $t^{\Gl_i}$, and accordingly to denote 
the structure constants $C_{\Gl_i\Gl_j\Gl_k}$ or $C_{\Gl\Gm\Gn}$, 
for $\Gl,\Gm,\Gn$ exponents. 
The expressions of the prepotentials for the various $ADE$ cases
have been listed in the literature (\refs{\No-\DFLZ} 
and further references therein). 

Now, by inspection, we observe the following
\smallskip
\nind {\bf Fact 1 :} {\sl For the $A_n$, $D_{2n}$, $E_6$ and $E_8$ cases, 
there exists a choice of flat coordinates for which all the 
coefficients of $F$ are real positive. For $D_{2n+1}$ and $E_7$ 
there is no such choice. }
\medskip
 Two remarks are in order. \par 
\nind First, why is it meaningful to
look at reality and positivity properties in a problem that 
looks intrinsically complex?\par
\nind Secondly it is curious that 
this splitting of the $ADE$ classification scheme into the same  
two sub-families appears also in other contexts. Let us quote 
\item{1)} the structure  of the modular invariant partition 
function of conformal field theories with a $\slh(2)$ current algebra.
The latter are known to follow an $ADE$ classification scheme
\refs{\CIZ,\Ka}. The question is to know if this partition function, 
which is  a certain sesquilinear form with  non negative integer
coefficients,  may or may not be written as a sum of %modulus squares
blocks
\eqna\Iza
$$\eqalignno{ Z & =\sum {\CN}_{\Gl\bar\Gl} \chi_{_{\Gl}} \bar\chi_{_{\bar\Gl}}
\qquad \qquad \CN_{\Gl\bar \Gl } \in \IN & \Iza a \cr
& \buildrel {?}\over= \sum_{i} |\sum_{\Gl\in \hT_\Ga} \chi_{_{\Gl}}|^2 
\ .& \Iza b \cr}$$
For example, the cases labelled by $D_{10}$ and $E_7$ read respectively
\eqna\Izb
$$\eqalignno{
 Z^{(D_{10})}&= |\chi_1+\chi_{17}|^2+|\chi_3+\chi_{15}|^2+|\chi_5+\chi_{13}|^2
+|\chi_7+\chi_{11}|^2+2|\chi_9|^2  & \Izb a\cr
Z^{(E_7)} &=|\chi_1+\chi_{17}|^2+|\chi_5+\chi_{13}|^2
+|\chi_7+\chi_{11}|^2+ |\chi_9|^2 
+\( (\chi_3+\chi_{15}) \chi_9^* + {\rm complex\ conj.}\) \cr
&= |\chi_1+\chi_{17}|^2+|\chi_5+\chi_{13}|^2
+|\chi_7+\chi_{11}|^2+ |\chi_9+\chi_3+\chi_{15}|^2 
-|\chi_3+\chi_{15}|^2 & \Izb b\cr }$$
(for more details and explanation of notations, see below sect. 5). 
\item{2)} the positivity of the structure constants of the 
``Pasquier algebras" to be discussed below;
\item{3)} the existence or non-existence of a ``flat connection" 
on the  path algebra on the Dynkin diagram \OEK;
\item{4)} the positivity of the coefficients of the prepotential
just discussed; 
\item{5)} the positivity properties of the coefficients of the 
 factors of the Poincar\'e polynomial of the local (or ``chiral") 
ring of the singularity. Let us discuss briefly this latest aspect, as it 
does not seem to be generally known. For any of the $ADE$ singularities, 
let $p$ denote the minimal number of non morsian variables $X_i$ that enter
the singular polynomial. Let us write the 
Poincar\'e polynomial in the form 
%$$ 
\eqn\Ia{P(t) =\prod_{i=1}^p{(1-t^{h-\deg(X_i))}\over (1-t^{\deg(X_i)})} }
% $$
%
in terms of the degrees of the variables $X_i$ and of the Coxeter 
number $h$, equal to the degree of the singular polynomial.  
\medskip
\halign{ # & # & # &  # & # \cr   
 \quad & \qquad $h$ & \qquad $p$ & \qquad $\{ \deg(X_i)\}$ & \qquad{\rm exponents}\quad $\Gl$ \cr
 \qquad \qquad $A_n$ & \qquad   $n+1$ &\qquad 1&\qquad $1$ &\qquad $1,2,\cdots
,n$ \cr
 \qquad \qquad $D_{\ell+2}$  &\qquad  $2(\ell+1)$ &  \qquad 2 
         &\qquad  $2,\ell$ &\qquad $1,3,\cdots,2\ell+1,\ell+1$ \cr
 \qquad \qquad $E_6$ & \qquad 12  & \qquad 2 & \qquad 3,4 &\qquad $1,4,5,7,8,11
$\cr
 \qquad \qquad $E_7$ & \qquad 18 & \qquad 2  & \qquad 4,6
& \qquad $1,5,7,9,11,13,17$  \cr
 \qquad \qquad $E_8$ & \qquad  30 & \qquad 2  & \qquad 6,10& \qquad 
$1,7,11,13,17,19,23,29$ \cr }
It is then an easy and amusing exercise to check that $P(t)$ 
may be written as a product of $p$  factors with positive coefficients 
only in the first subfamily. 
 (I owe this observation to M. Bauer \MB). %[unpublished]).
This is somehow the multiplicative counterpart of the property 1) 
mentionned above. 

The interesting thing is that the simultaneous occurrence 
of several of these properties seem to extend beyond the $ADE$
case discussed here. 
The status of these various occurences is however not the same. 
I think it is fair to say that 1) is the best understood, 
as it is related to a structural property of the underlying conformal
field theory. 2) is related to 4) as we shall see soon, but 
I doubt that 4) may be extended beyond the case of 
simple singularities, as the prepotential is then no longer a
polynomial. Finally it seems that 5) does not generalize: 
for some singularities believed to be  in correspondence with 
some conformal field theory, property 5) may fail while 1) and 
2) are true (for example, the singularity associated with the 
fusion potential of $\slh(4)_4$). 

\medskip
In fact we are not going to make use of Fact 1 for generic $\bt$, 
but only for a particular case, obtained by the so-called \Che 
specialization. This refers to the deformation of the $ADE$ 
singularity for which all the flat coordinates but the one, $t^n$, with the 
smallest degree $\deg(t^n)=2$, 
(the largest exponent $\Gl_n=h-1$), i.e. the ``less relevant" 
in the language of physics, is kept non zero. As it is the only 
parameter in the homogeneous deformed polynomial $W(X_1,\cdots,X_p, t^n)$, 
one may rescale it to $t^n=1$. 
The origin of the denomination is that for the $A_n$ case, 
the deformed polynomial reads then $W_{A_n}(X_1;t^n=1)= T_{n+1}(X_1)$%
%+X_2^2+X_3^2$
, with $T_{n+1}(x)$ the degree $n+1$ \Che polynomial of first kind, 
$T_{n+1}(x)=2 \cos (n+1)\theta$ if $x=2\cos \theta$.

\medskip
We also need some notations on the $ADE$ Dynkin diagrams. Let $G_{ab}$
denote the adjacency matrix of the Dynkin diagram under consideration: 
$a,b=1, \cdots, n$  label the vertices. The corresponding Cartan matrix 
is $C_{ab}= 2\Gd_{ab} -G_{ab}$. The eigenvectors $\psi^{(\Gl)}$ 
and eigenvalues
of these symmetric matrices are indexed by the Coxeter exponents $\Gl$, 
 %a set of $n$ integers between $1$ and $h-1$, $h$ the Coxeter number
%$$
\eqn\Ib{ G_{ab} \psi^{(\Gl)}_b = 2 \cos{\pi \Gl \over h} \psi^{(\Gl)}_a\ .}
%$$
%
The $\psi^{(\Gl)}$ may be chosen orthonormal. 

\medskip 
Then we can state the \par
\smallskip
\nind {\bf Fact 2 : } {\sl The structure constants of the chiral ring in 
the \Che specialization are diagonalized by the $\psi^{(\Gl)}$}
%$$
\eqn\Ic{ M_{\Gl\Gm}^{\ \ \Gn}:= C_{\Gl\Gm}^{\ \ \Gn}(t^n=1)=
\sum_a {\psi^{(\Gl)}_a\psi^{(\Gm)}_a\psi^{(\Gn)*}_a\over \psi^{(1)}_a}}
%$$
%
\smallskip
Here I have introduced the notation $M$ to be used in the 
forthcoming discussion. In the denominator of the right hand side, 
there appears the exponent $1$, that yields the largest eigenvalue of 
the matrix $G$. By the Perron-Frobenius theorem, all the components 
of $\psi^{(1)}$ are non vanishing and of the same sign. 

Fact 2 is not a surprise in the $A_n$ case, where it follows from the 
combined work of Verlinde \EV\ and Gepner \DGe. Indeed the above 
structure constants reduce then to the fusion coefficients of the 
$\slh(2)$ algebra, for a value of the level (central extension)
equal to $k=n-1$, and the latter are known to have an interpretation 
in terms of the chiral ring of  a topological field theory. 
 For the other $D$ and $E$ cases, the observation 
was made (in essence, not quite in these terms) by Lerche and 
Warner \LW, and made more systematic and extended in \DFLZ. 

\bigskip
The previous formula suggests to consider also the dual
algebra (we shall see below that the word ``dual" is legitimate), 
with structure constants
%$$ 
\eqn\Id{N_{ab}^{\ \ c}:=
\sum_\Gl {\psi^{(\Gl)}_a\psi^{(\Gl)}_b\psi^{(\Gl)*}_c\over \psi^{(\Gl)}_1}}
%$$
%
where the sum runs over the exponents $\Gl$ of the case at hand. 
This definition depends on a choice of a vertex denoted 1 for which 
all the $\psi^{(\Gl)}_1$ are non-vanishing. Such a vertex exists  
for all the $ADE$ cases. There may remain, however, some arbitrariness
in the choice of that vertex 1 and also, in the case $D_{2n}$ for 
which an exponent occurs with multiplicity 2, in the choice of the 
basis $\psi^{(\Gl)}_a$. Now comes the
\medskip
\nind {\bf Fact 3 : } {\sl For the $A_n$, $D_{2n}$, $E_6$ and $E_8$ cases, 
 there exists a choice of vertex 1 and of 
the basis $\psi^{(\Gl)}_a$ such that the structure constants 
$M_{\Gl\Gm}^{\ \ \Gn}$ and $N_{ab}^{\ \ c}$ are all non negative. 
For the cases $D_{2n+1}$ and $E_7$, there exists no such choice.}
\smallskip
 
Note that the non-negativity of the $M$ is a simple consequence 
of Fact 1 $\cap$ Fact 2. For the $ADE$ cases, the numbers $N$ turn 
out to be integers (with an adequate choice of 1 and the basis). 
The interpretation of these numbers in the various contexts in 
which they occur (conformal field theories, topological theories
and singularities, lattice models) has remained elusive so far. 
In contrast, the $M$ that are in general non integers but rather 
algebraic  numbers, have such an interpretation: 
they give the structure constants of the chiral ring of the \Che 
specialization, as just explained; in the context of conformal field 
theories and integrable lattice models, they give the coupling 
constants of field operators \Pas, \EV, \PZun. It is in that context
that this algebra was first introduced by Pasquier \Pas, whence the 
name of Pasquier algebras that I give to the pair of $M$ and $N$
 algebras. 

%%%%%%%%%%%%%%%%%%%%%%%%%%%%%%%%%%%%%%%%%%%%%%%%%%%%%%%%%%%%%%%%

\newsec{C-algebras}
\subsec{Definitions and examples}
\nind The appropriate language to discuss these Pasquier algebras
is that of {\it C-algebras}, (``C" for character), introduced in the 
40's by Kawada and recently reviewed and revived by Bannai and
Ito \BI.

\smallskip
\nind {\bf Definition : }
{\sl An algebra $\gA$ over $\IC$ with a given basis $x_1, \cdots, x_n$, 
is a C-algebra if it satisfies the following axioms:
\item{i)} it is a commutative and associative algebra with {\rm real}
structure
constants $p_{ab}^{\ \ c}$, i.e. $x_a. x_b =\sum_c p_{ab}^{\ \ c} x_c$;
%\foot{Here and in the rest of this paper, summation over repeated 
%indices is understood, unless explicitly stated otherwise.}
\item{ii)} it has an identity element, denoted $x_1$, i.e. 
$p_{1a}^{\ \ b}=\Gd_{ab}$;
\item{iii)} there is an involution on the generators $x_a \mapsto
x_{\bar a}$ that is an automorphism of the algebra, i.e.
$p_{ab}^{\ \ c}=p_{\bar a\bar b}^{\ \ \bar c}$;
\item{iv)} $p_{ab}^{\ \ 1}= k_a \Gd_{a\bar b}$, with $k_a$ a real 
positive number $k_a>0$;
\item{v)} the $k_a$ form a one-dimensional representation of the algebra.} 

Among the various consequences of these axioms, is the fact that 
$\gA$ is semi-simple. There are $n$ one-dimensional representations
of the algebra, that we label by an index $\Gl$ taking $n$ values : 
$x_a \mapsto p_a(\Gl)\in \IC$. The value $\Gl=1$ refers to the 
special representation of axiom v): $p_a(1)=k_a$. If
$e_\Gl$ denote the corresponding 
idempotents, one may decompose $x_a=\sum_\Gl p_a(\Gl) e_\Gl$ . The matrix 
$p_a(\Gl)$ is invertible, let $q_\Gl(a)$ denote the  matrix
such that $\sum_\Gl p_a(\Gl) q_\Gl(b)= \kappa \Gd_{ab}$, 
$\kappa := \sum_a k_a$. More explicitly, 
the matrices $P_a$ of elements $(P_a)_b^{\ c} =\sqrt{{k_c\over k_b}}
p_{ab}^{\ \ c}$ form a representation of the algebra $\gA$. They 
are normal and commuting,  and thus diagonalizable in a common
 orthonormal basis $\psi^{(\Gl)}_a$. All $\psi_1^{(\Gl)}$ and
$\psi^{(1)}_a$ are non vanishing and may thus be chosen real positive.
One may write
\eqn\IIa{
\eqalign{
p_{ab}^{\ \ c}&=\sqrt{{k_a k_b\over k_c}} \sum_\Gl 
{\psi^{(\Gl)}_a \psi^{(\Gl)}_b \psi^{(\Gl)\,*}_c \over \psi^{(\Gl)}_1} \cr
\sqrt{k_a} &= {\psi^{(1)}_a\over \psi^{(1)}_1}  %& \IIa 
\cr
p_a(\Gl)&= { \psi^{(\Gl)}_a \psi^{(1)}_a\over \psi^{(\Gl)}_1\psi^{(1)}_1}\cr
q_\Gl(a)&={\psi^{(\Gl)\, *}_a\psi^{(\Gl)}_1\over\psi^{(1)}_a\psi^{(1)}_1}\cr
}}
and let  $\hat k_\Gl$ be such that 
%$$ 
\eqn\IIb{\sqrt{\hat k_\Gl} = {\psi^{(\Gl)}_1\over \psi^{(1)}_1}\ .}
% $$ 
%
One may then show that the dual  $\hat \gA$ of
$\gA$, defined as the set of linear maps from $\gA$ into $\IC$, is
endowed with a structure of C-algebra: its basis is labelled by the $\Gl$, 
its one dimensional representations are provided by  the $q_\Gl(a)$, 
among which $q_\Gl(1)= \hat k_\Gl$ are positive, 
and the structure constants of the algebra are 
%$$ 
\eqn\IIc{
q_{\Gl\Gm}^{\ \ \Gn}=
\sqrt{{\hat k_\Gl \hat k_\Gm\over \hat k_\Gn}} \sum_a 
{\psi^{(\Gl)}_a \psi^{(\Gm)}_a \psi^{(\Gn)\,*}_a \over \psi^{(1)}_a} \ .}
%$$
%
The $k_a$ and $\hat k_\Gl$ are called the Krein parameters of the algebras.
They satisfy $\kappa=\sum_a k_a=\sum_\Gl \hat k_\Gl=1/\psi_1^{(1)\, 2}$.
 
Alternatively, one may regard this dual  $\hat \gA$ 
 as a second C-algebra structure
on $\gA$, with basis $ \kappa e_\Gl$ and idempotents $x_a$. To recapitulate, 
 $\gA$ is endowed with a  pair of dual C-algebra structures, 
one with multiplication $.$,
 structure constants $p_{ab}^{\ \ c}$  in the  basis $x_a$, 
and idempotents $e_\Gl$, and the other with multiplication 
$\circ$,  structure constants $q_{\Gl\Gm}^{\ \ \Gn}$  in the  basis 
$\kappa e_\Gl$ and idempotents $x_a$
%
%
%$$
\eqn\IId{\eqalign{ x_a.x_b &= \sum_c p_{ab}^{\ \ c} x_c\ ,\qquad e_\Gl.e_\Gm = 
\Gd_{\Gl\Gm} e_\Gl \cr
 \kappa e_\Gl \circ \kappa e_\Gm & = \sum_\Gn q_{\Gl\Gm}^{\ \ \Gn}\, \kappa e_\Gn\ , 
\qquad x_a\circ x_b  = \Gd_{ab} x_a \ . \cr }}
%$$ 
%

\medskip
\nind {\it Examples}: \par 
\nind 1. Character and class algebras of a finite group. Let $\GC$
be a finite group, $C_a$ denote its equivalence classes, 
$(\Gr)$ its irreducible representations, $\chi^{(\Gr)}$ their
characters, $\chi^{(\Gr)}_a$ the value of these characters on class $a$; 
$a=1$ refers to the class of the identity, 
$\Gr=1$ to the identity representation; $d_\Gr=\chi^{(\Gr)}_1$ is the 
dimension of representation $\Gr$. One has two dual algebras
%$$ 
\eqn\IIe{\eqalign{C_a C_b &= C_{ab}^{\ \ c} C_c \cr
\chi^{(\Gl)}\chi^{(\Gm)} &= K^{\Gl\Gm}_{\ \ \Gn} \chi^{(\Gn)}\ . \cr }}
%$$
%
Introducing the ${\widehat \chi}_a^{\Gl}= \sqrt{{|C_a|\over |\GC|}}
\chi_a^{(\Gl)} $, orthonormal by virtue of the standard 
orthogonality and completeness relations of characters, one may 
write
%$$
\eqn\IIf{
\eqalign{
p_{ab}^{\ \ c}= C_{ab}^{\ \ c} &= \sqrt{{|C_a||C_b|\over |C_c|}}
\sum_\Gl {\chih^{(\Gl)}_a\chih^{(\Gl)}_b\chih^{(\Gl)*}_c\over \chih^{(\Gl)}_1}
\cr
q_{\Gl\Gm}^{\ \ \Gn}={d_{\Gl}d_{\Gm}\over d_{\Gn}}
K^{\Gl\Gm}_{\ \ \Gn} &= {d_{\Gl}d_{\Gm}\over d_{\Gn}}
\sum_a{\chih^{(\Gl)}_a\chih^{(\Gm)}_a\chih^{(\Gn)*}_a\over\chih^{(1)}_a}\cr}}
%$$
%
The two dual algebras have integer Krein parameters $k_a=|C_a|$, 
$\hat k_\Gl=d_\Gl^2$ with  the well known relation $|\GC|=\sum k_a=
\sum\hat k_\Gl=\sum d_\Gl^2$. 

\medskip
\nind 2. The Pasquier algebras  introduced above are obviously a 
pair of dual C-algebras. The structure constants 
$p_{ab}^{\ \ c}$ and $q_{\Gl\Gm}^{\ \ \Gn}$  are 
respectively proportional to  $N_{ab}^{\ \ c}$
and $M_{\Gl\Gm}^{\ \ \Gn}$, as indicated in \IIa\ and \IIc.
In that case, in contrast with example 1, the Krein parameters
are not integers. Among these Pasquier algebras, there are the
fusion algebras of affine algebras $\ggh$. In that case, the
two dual algebras are in fact isomorphic: this is due to the fact
that according to the Verlinde formula, the diagonalizing matrix
is the {\it symmetric} unitary matrix $S$ of modular transformations
of the affine characters \EV. Also, in that case, the Krein 
parameters are equal to $\hat k_\Gl= \({S_{1\Gl}\over S_{11}}\)^2$, 
that is $\hat k_\Gl= D_\Gl^2$, the square of the quantum dimension 
of the corresponding representation of $\ggh$. This is thus a 
 quantum deformation of the finite group situation of the previous example.  

\par

%\bigskip
\subsec{C-subalgebras}
\nind One then defines C-subalgebras of a C-algebra : 

\nind 
{\bf Definition :} {\sl 
Given a C-algebra with a basis $\{x_a\}$, $a=1, \cdots,n$,
a C-subalgebra is a C-algebra generated by a subset of the $x_a$, 
$a\in T$, $T\subset \{1, \cdots, n\}$, closed under multiplication, i.e.
if $a,b\in T$, $p_{ab}^{\ \ c}\ne 0$ only if $c \in T$. }

Note that this condition implies that $T$ contains $1$ and is stable 
under the involution $a \mapsto  \bar a$ \BI. 

We shall be mainly interested in the situation where the two dual
algebras have non negative structure constants. Then there is a 
remarkable theorem that tells us that the existence of a C-subalgebra
in $\gA$ implies the existence of a C-subalgebra in the dual.
More precisely, suppose $\gA$ has a C-subalgebra $\gA_T$ associated with 
a subset $T$. One may then define an equivalence relation 
$a \sim b$ if $\exists c\in T\ : \ p_{ac}^{\ \ b}\ne 0$, and  
 there is a partition of the set $\{1,2, \cdots n\}$
into equivalence classes, $T_i$, $i=1, \cdots, p$, $T_1\equiv T$. 
Let $\Gr= \sum_{a\in T} k_a$ and let $X_i:= \sum_{a\in T_i} x_a$. 
One also defines the subset $\hat T$ of the dual basis  by 
the decomposition of $X_1=\sum_T x_a$
into idempotents $X_1=\Gr \sum_{\Gl\in \hat T} e_\Gl$. 

\smallskip
\nind {\bf Theorem} (Bannai-Ito \BI, theorem 9.9): {\sl
Consider a C-algebra $\gA$ with non negative structure constants 
$p_{ab}^{\ \ c}$ and $q_{\Gl\Gm}^{\ \ \Gn}$. With the notations 
just introduced, 
\item{i)} the ${1\over \Gr}X_i$, $i=1,\cdots, p$, 
 generate themselves a C-algebra, called the 
quotient C-algebra $\gA/\gA_T$, with  a product inherited from $\gA$;
\item{ii)} the $\kappa e_\Gl$, for $\Gl\in \hat T$, generate  
a C-subalgebra $\hat\gA_{\hat T}$ of the  dual algebra $\hat \gA$;
\item{iii)} these two C-algebras are dual to one another. }

Thus one has a dual pattern of subsets $T$ and $\hat T$, 
of C-subalgebras $\gA_T$ and $\hat\gA_{\hat T}$, and of quotients
$\gA/\gA_T$ and $\hat\gA/\hat\gA_{\hat T}$
with the isomorphisms $\widehat{\gA/\gA_T}\cong \hat\gA_{\hat T}$ and 
vice versa $\hat\gA/\hat\gA_{\hat T}\cong \widehat{\gA_T}$. 

One proves also that all $X_i$ may be expanded on the $e_\Gl$, 
$\Gl\in \hat T$, and conversely. Recalling that $x_a=\sum_\Gl p_a(\Gl) e_\Gl$
and $\kappa e_\Gl=\sum_a q_\Gl(a) x_a$, with expressions 
of $p_a(\Gl)$ and $q_\Gl(a)$ given in \IIa, we find that
\eqn\IIg
{\eqalign{\sum_{a\in T_i} p_a(\Gl) &=0 \quad {\rm if}\ \Gl\notin \hat T \cr
{\rm thus }\quad \sum_{a \in T_i}   \psi^{(\Gl)}_a \psi^{(1)}_a &=0 
\quad {\rm if}\ \Gl\notin \hat T \cr
{\rm for\  }\Gl \in \hat T\quad 
q_\Gl(a) &= {\psi_a^{(\Gl)\, *}\over \psi^{(1)}_a}
{\psi_1^{(\Gl)}\over \psi^{(1)}_1}
 {\rm 
\quad  independent \ of\ } a\in T_i \ .\cr
}}
These two conditions may be conveniently assembled into a single one
\eqn\IIh{
\forall \Gl , \forall T_i, \forall a\in T_i\qquad
\sum_{b\in T_i} \psi_b^{(\Gl)} \psi_b^{(1)} 
= \Gd_{\Gl\in \hat T} {\psi_a^{(\Gl)}\over \psi_a^{(1)}}
\sum_{b\in T_i} (\psi_b^{(1)})^2 \ , }
a form that will be useful in the sequel. 
It is also easy to  write explicitly the expressions of the 
structure constants of the quotient algebras. For example, 
from $X_i=\sum_{a\in T_i} x_a$ it follows that ${1\over \Gr}X_i.
{1\over \Gr}X_j=\sum_k \bp_{ij}^{\ \ k}\,{1\over \Gr}X_k$
with 
\eqn\IIi{\bp_{ij}^{\ \ k} ={1\over \Gr}
 \sum_{c \in T_k} p_{ab}^{\ \ c}\ , \ \forall a\in T_i, b\in T_j\ . }
\medskip
In the following two sections, 
I shall present two applications of this theorem. % on C-subalgebras.
The first deals with reflection groups and their folding, the second 
with conformal field theories. The first starts with C-subalgebras 
of the $M$ algebra (subject to an additional constraint), 
the second with those of the $N$ algebra. 
%For the sake of simplicity they  will both be presented in the 
%context of $ADE$ Dynkin diagrams, i.e. of finite (Coxeter) reflection
%groups on the one hand, and of $\slh(2)$ conformal field theories
%on the other. It should be stressed, however, that they %may be 
%extend to a much larger class of diagrams and cft's. See 
%my lecture  at the Taniguchi Symposium \Tani\
%for the reflection groups 
%and \PZde\ for the discussion of cft's. 

%\medskip
%%%%%%%%%%%%%%%%%%%%%%%%%%%%%%%%%%%%%%%%%%%%%%%%%%%%%%%%%%%%%%%%

% 
%\medskip
\newsec{Folding of $ADE$ Dynkin diagrams} 
\subsec{The problem}
\nind It is well known that non simply laced Dynkin diagrams 
(of type $B_n,C_n,F_4,G_2$) may be 
obtained by folding the simply laced ones% [who?]
, using the symmetries of the original diagram. The extension to
Coxeter diagrams of $H$ or $I$ type, associated with the 
non-crystallographic Coxeter groups, seems more recent \refs{\Yano,\fold}.
%It has been rediscovered recently 
%in the context of topological field theories [Dubrovin, Z]. 
In all these works, one is given a simply laced Dynkin %Coxeter graph 
diagram describing the scalar products of a set of simple roots $\{\Ga_a\}$, 
$a=1, \cdots, n$, according to
\eqn\IVaa{(\Ga_a,\Ga_b)=2\Gd_{ab}-G_{ab}}
($G$ the adjacency matrix as in \Ib). Then 
  a certain partition is found of this set into subsets 
$\{\Ga_a, a\in T_i\}$ of mutually orthogonal roots
\eqn\IVa{ (\Ga_a,\Ga_b)=0 \qquad {\rm if \quad } a, b \in T_i\ .}
Let $S_a$ denote the reflection in the hyperplane orthogonal to 
$\Ga_a$ through the origin, and $G$ the group generated by all the 
$S_a$, $a=1,\cdots, n$. 
{}Then one forms the products 
\eqn\IVb{ R_i =\prod_{a\in T_i} S_a}
in which the order is immaterial, since the $\Ga$ are orthogonal
within the same $T_i$, and thus the $S_a$ commute. 
The group $G'$ generated by the $R_i$ is clearly a subgroup of 
$G$. Since $G$ is a Coxeter group (of finite order), $G'$ is also
of finite order, hence in the $A-I$ list. The corresponding Coxeter
diagram thus results from identifying the vertices of a same
block $T_i$, while the superscript of an edge $i-j$, which 
yields the order of the element $R_i R_j$ may be 
computed easily in terms of the original $S_a$. 
One finds 
empirically the adequate foldings of the $A,D,E$ diagrams 
necessary to manufacture all the others (see Fig. 1). 
For example, the order 5 of the product $R_2 R_3$ in  the 
diagram $H_3$, {\it i.e.} the smallest power $m$ s.t. 
$(S_2 S_3 S_4 S_6)^m= I$ is simply the order of
the Coxeter element of the $A_4$ Coxeter group generated by these
four reflections.

As far as I can see, this procedure is, however, empiric, and
doesn't say which folding does the job and in which subspace of the 
original $n$-dimensional space the subgroup acts. 
In the fairly different context of topological field theories (tft), 
a parallel observation was made. Starting from the so-called
minimal tft's labelled by  $ADE$ Dynkin diagrams, i.e. solutions
of the WDVV equations of the type mentionned in sect. 2, one finds that 
there are other solutions obtained by restriction of the latter. 
In such a restriction, only a subset of the flat coordinates $t$ is 
kept non-vanishing. These non-vanishing $t$'s are labeled by the 
Coxeter exponents of some %, now labelled by 
non simply laced Coxeter groups \refs{\Dub, \Zdub}. These 
restrictions are consistent with the algebra of the tft, in the sense
that they correspond to a sub-algebra of the $C_{\Gl\Gm}^{\ \ \Gn}(\bt)$. 
If we consider the \Che specialization and recall Fact 2 of sect. 2, 
this means that the Pasquier algebra $M$ of the original $ADE$ diagram
admits a sub-algebra, whose generators are labelled by the exponents of a
Coxeter group of type $B,C,F,G-I$ \Zdub. 

In fact there is a strong connection between the two observations, 
and through the theory of C-algebras, one is able to
answer the previous objection and {\it determine} the folding 
through the study of the subalgebras of $M$ type. 

%%%%%%%%%%%%%%%%%%%%%%%%%%%%%%%%%%%%%%%%%%%%%%%%%%%%%%%%%%%%%%%%%%

\subsec{From a $M$-subalgebra to a subgroup}

\noindent
Consider a simply laced $ADE$ Dynkin diagram such that 
the structure constants $M$ and $N$
are non negative (see Fact 3 of sect. 2). 
Recall that all Dynkin diagrams may be 2-coloured, {\it i.e.} their 
vertices may be assignend a $\IZ_2$ grading $\tau$, the ``colour",  such that 
$G_{ab}=0$ if $\tau(a)=\tau(b)$. Now suppose that a subalgebra of the $M$
algebra has been found, {\it i.e.} a subset $\hat T$ of exponents
such that 
\eqn\IVIaa{\Gl,\Gm\in \hat T\qquad
 M_{\Gl\Gm}^{\ \ \Gn}\ne 0 \Rightarrow\, \Gn\in \hat T\ ; }
 the subset $\hat T$ of exponents is assumed to be stable under 
%the action of $\Gs\ :
$\ \Gl\mapsto h-\Gl$. 
The positivity condition tells us that we
are in the conditions of the theorem of sect. 3.2. 
Because here we start from a C-subalgebra of the $M$ (or $q$)
algebra, the theorem has to be transposed to its 
dual version,     namely 
\item{(i)} there is a partition of the set of exponents into equivalence 
classes $\hat T_\Ga$, 
\eqn\IVIab{\Gm \sim \Gn \quad {\rm if} \quad \exists \Gl\in \hat T\, ,\
M_{\Gl\Gm}^{\ \ \Gn}\ne 0\ ; }
\item{(ii)} there exists a special subset $T$ of
the dual set of vertices that contains $1$; %, such that
%% 
%\eqn\IVIac{ \sum_{\Gm\in \hT_\Ga^} \psi_a^{(\Gm)} \psi_1^{(\Gm)}
%= \Gd_{a\in T} {\psi_a^{(\Gl)}\over \psi_1^{(\Gl)}}
%\sum_{\Gm\in \hT_\Ga} (\psi_1^{(\Gm)})^2 \qquad\quad \forall\Gl\in\hT_\Ga\ ,}
%%
%so that clearly $1\in T$;
\item{(iii)} the set $T$ enables one to define 	a dual equivalence
relation: $b\sim c$ if $\exists a\in T$ such that $N_{ab}^{\ \ c }\ne 0$, 
and hence a partition of the set of vertices into equivalence classes
$T_i$;  
\item{(iv)} the relation  \IIh\ is satisfied. 

%\noindent
%In words, \IVIac\ says that the ratio ${\psi_a^{(\Gl)}/ \psi_1^{(\Gl)}}$
%is independent of $\Gl\in T_\Ga^*$ if $a\in T$.
%There is an analogous dual statement (see 
%\DFZ, (formula (4.36)) or \DF, (8.31))
%
%\eqn\IVIb{ \sum_{b\in T_i} \psi_b^{(1)}\psi_b^{(\Gl)}
%= \Gd_{\Gl\in \hat T} {\psi_a^{(\Gl)}\over \psi_a^{(1)}}
%\sum_{b\in T_i} (\psi_b^{(1)})^2 \qquad\quad \forall a \in T_i \ .}

\medskip
Now the assumption made above that $\hat T$ is stable under $\Gl
\mapsto h-\Gl $
implies that: (i) the same is true for each class $\hat T_\Ga\,$;
(ii) the class $T$ contains only vertices $a$ satisfying 
$\tau(a)=\tau(1)$; % (which may be  taken  equal to zero by convention); 
(iii) more generally all the vertices within a same 
class $T_i$ have the same colour $\tau$ and thus the corresponding 
roots are mutually orthogonal. These are trivial consequences of the
symmetry of %action of $\Gs$ on 
the $\psi$
\eqn\IVIba{\psi_a^{(h-\Gl)} =(-1)^{\tau(a)}  \psi_a^{(\Gl)}}

\bigskip
I now claim that with this pattern of subalgebras one may associate
a subgroup of $G$; it is again described by a graph, whose 
vertices are in one-to-one correspondence with the classes $T_i$
and whose set of exponents is $\hat T$.
This subgroup is generated by reflections in the hyperplanes orthogonal
to some $\Gb$, that are 
 some linear (real) combinations of the roots $\Ga$: 
\eqn\IVIc{ \Gb_i= \CN_i \sum_{a\in T_i} \psi^{(1)}_a \Ga_a \  ;}
the normalisation 
is adjusted so that $(\Gb_i,\Gb_i)=2$, namely
\eqn\IVId{  \CN_i^2 \sum_{a\in T_i} (\psi^{(1)}_a)^2 =1 \  }
(since the $\Ga_a$, $a\in T_i$ are mutually orthogonal).
One verifies, using \IIh, that the product 
$\prod_{a\in T_i} S_a$ has the same action as 
the reflection $R_i$ in the hyperplane orthogonal to $\Gb_i$, 
in the subspace spanned by the $\Gb$ \Tani. 

\medskip
%%%%%%%%%%%%%%%%%%%%%%%%%%%%%%%%%%%%%%%%%%%%%%%%%%%%%%%%%%%%%%%%%%%%%%
%
%\subsec{The folded metric }
%\noindent
The scalar products of two distinct roots $\Gb_i$  and $\Gb_j$ 
is non positive, as follows from the same property for the original
simple positive roots $\Ga_a$ and from the positivity of the 
$\psi^{(1)}_a$
$$ (\Gb_i,\Gb_j)= \CN_i\CN_j \sum_{a\in T_i \atop b\in T_j} 
(\Ga_a,\Gb_b) \psi^{(1)}_a \psi^{(1)}_b \le 0 \ .$$
The metric defined on the original roots may be diagonalized 
by the $\psi$
\eqn\IVIk{ 
g_{ab}=(\Ga_a,\Ga_b)=
 \sum_{ {\rm exponents}\ \Gl} g^{(\Gl)}\psi_a^{(\Gl)}\psi_b^{(\Gl)\, *}\ ,}
with $g^{(\Gl)}=2-2\cos {\pi\Gl\over h}$. 
{}From the expressions of the new roots $\Gb_i$ it is easy to compute
the new metric, making use again of \IIh
\eqn\IVIm{\eqalign{
g_{ij}&=(\Gb_i,\Gb_j)=\CN_i^{-1} \CN_j^{-1} \sum_{\Gl\in \hat T}
g^{(\Gl)} {\psi^{(\Gl)}_a\over \psi^{(1)}_a} 
{\psi^{(\Gl)\, *}_b\over \psi^{(1)}_b}
\quad \forall \, a \in  T_i \ , \forall \, b \in  T_j 
 \cr 
&= \sum_{\Gl\in \hat T}
g^{(\Gl)} \Psi^{(\Gl)}_i \Psi^{(\Gl)\, *}_j \cr }}
in terms of the new eigenvectors
\eqn\IVIn{\eqalign{
\Gl\in \hat T \qquad 
\Psi^{(\Gl)}_i &= {\CN}_i^{-1} {\psi^{(\Gl)}_a\over \psi^{(1)}_a}
\quad \forall \, a \in  T_i  \cr
& = \CN_i \sum_{a\in T_i} \psi^{(\Gl)}_a\psi^{(1)}_a \ .  \cr}}
These eigenvectors form an orthonormal system of rank $|\hat T|$. 
%%%%%%%%%%%%%%%%%%%%%%%%%%%%%%%%%%%%%%%%%%%%%%%%%%%%%%%%%%%%%%%%%%%%%%%%
\bigskip
\fig{The folding of ADE Dynkin diagrams of positive type.
Classes $T_i$ of vertices encompass  nodes on the same vertical.}
%{fol.eps}{9cm}\figlabel\foldi
{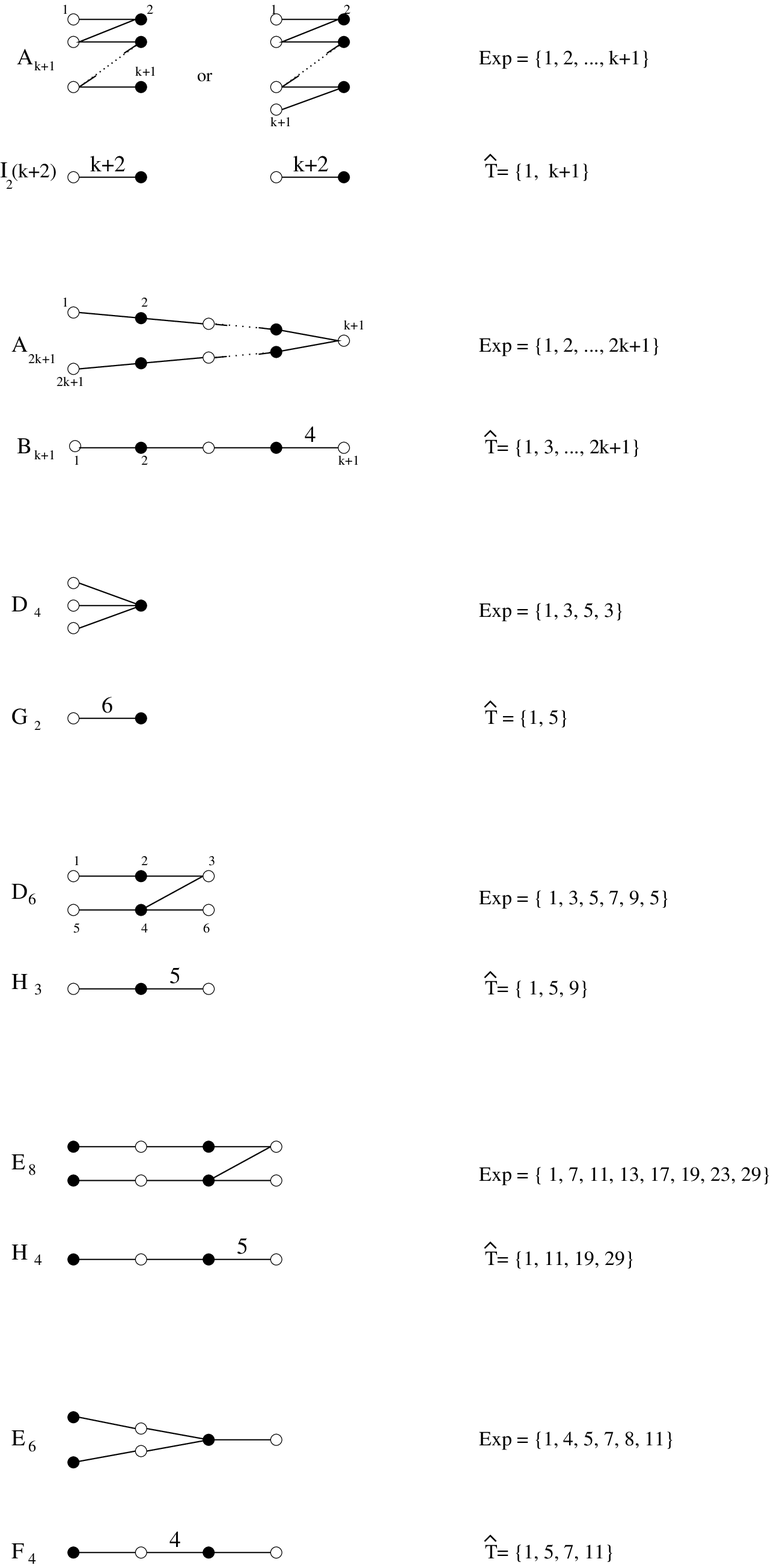}{95mm}\figlabel\foldi
%\fig{cont'd}
%{fo2.eps}{10cm}\figlabel\foldo

%\bigskip
\subsec{Discussion}
\noindent 
The reader may wonder what happens in \IVIc\ if the Perron-Frobenius
eigenvector $\psi^{(1)}$ is changed into another eigenvector. 
In fact, this has the effect of giving roots of the folded 
diagram that are simple but not positive.

The result of the procedure is presented  in fig. \foldi.
For each simply laced Dynkin diagram of type $A$, $D_{2\ell}$, 
$E_6$ or $E_8$, a systematic search of subalgebra of the $M$
algebra, satisfying the invariance of $\hat T$ under $\Gl\mapsto 
h-\Gl$ has been carried out. All  cases are not exhibited
in the Figures, as there is some redundancy. For example, any 
diagram of the previous type admits a subalgebra associated 
with $\hat T=\{1, h-1\}$. This corresponds to folding all 
vertices of a given colour onto one another, resulting in a 
2-vertex graph of type $I_2(h)$. This has been represented only 
for $A_{k+1} \mapsto I_2(k+2)$ or $D_4 \mapsto G_2\equiv I_2(6)$. 

\fig{A case of folding which is discarded by the assumption 
of positivity }{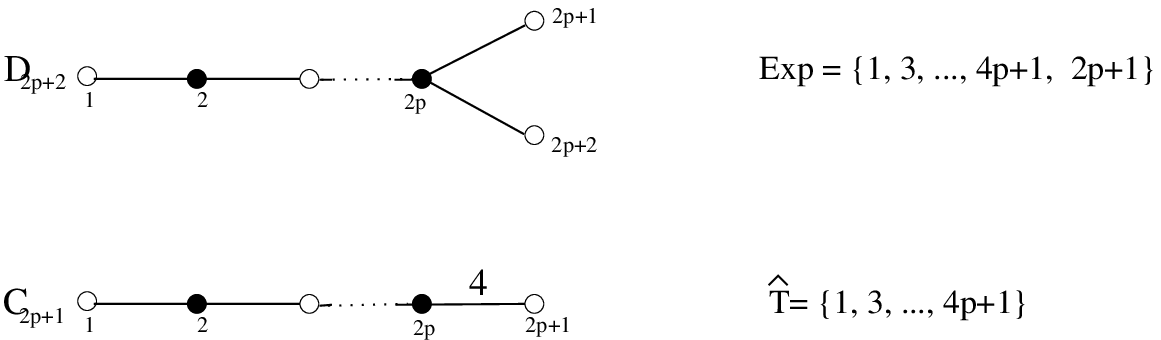}{10cm}\figlabel\foldu

By inspection of fig. \foldi,  the reader may convince herself
or himself that the procedure is exhaustive, in the sense that 
all non simply laced Coxeter diagrams, or all Coxeter groups, have 
been obtained. In fact, one possible folding of 
$D_{2p+2}$ into $C_{2p+1}$ (fig. \foldu) does not appear in the present
discussion. To expose the corresponding $M$ subalgebra of the 
$D_{2p+2}$ diagram requires indeed to change the basis of 
eigenvectors $\psi$ into another one, in which positivity is 
lost \PZun. In the present case, because of the isomorphism 
of $B_n$ and $C_n$ Coxeter-Weyl groups, this does  not hinder
the exhaustivity, 
but we may expect that the extension of the method to more general
cases may require relaxing the hypothesis of positivity.
We refer the reader to \Tani\ for a
 discussion of the appropriate extension of the 
present method.

%%%%%%%%%%%%%%%%%%%%%%%%%%%%%%%%%%%%%%%%%%%%%%%%%%%%%%%%%%%%%%%%
\newsec{Dynkin diagrams and RCFT}
\nind I shall be more concise on this part as it has already been 
expounded elsewhere \refs{\PZun,\PZde}. As recalled above in sect 
2, conformal field theories with a $\slh(2)$ current algebra have 
been classified according to an $ADE$ scheme. 
This manifests itself first in the form of their 
modular invariant genus 1 partition function, written as 
a sesquilinear form of characters $\chi_\Gl(q)$, $q=e^{2i\pi\tau}$,
of the affine $\slh(2)$
algebra at a given level $k$, with the integrable  weights $\Gl$
labelled by integers $1 \le \Gl \le k+1$. One proves \refs{\CIZ-\Ka} that 
 the possible expressions of that partition function 
\eqn\Va{Z= \sum {\CN}_{\Gl\bar\Gl} \chi_\Gl(q) \bar\chi_{\bar\Gl}(\bar q) 
\qquad \CN_{\Gl\bar \Gl}\in \IN}
are such that the {\it diagonal } terms $\Gl=\bar\Gl$ 
are the Coxeter exponents 
of one of the $ADE$ Dynkin diagrams of Coxeter number $h=k+2$. 
%A further analysis shows that more information about the Dynkin 
%diagram is encoded in the structure constants of the operator
%product expansion......

As alluded to in sect 2, the $A$, $D_{2\ell}$, $E_6$ and $E_8$ cases 
--and only those-- are such that $Z$ is a sum of blocks
$Z= \sum_\Ga |\sum_{\Gl\in \hT_\Ga} \chi_{_{\Gl}}|^2 $: 
\eqn\Vb{\Eqalign{
Z^{(A_n)} &= \sum_{\Gl=1}^n |\chi_\Gl|^2  &&&  k+2&=n+1 , \cr
Z^{(D_{2\ell})} &= \sum_{\Gl=1,3, \cdots, 2\ell-3} |\chi_\Gl+
\chi_{4\ell-2-\Gl}|^2  +2|\chi_{2\ell-1}|^2 &&& k+2&=4\ell-2 \cr
Z^{(E_6)} &= |\chi_1+\chi_7|^2+|\chi_4+\chi_8|^2+|\chi_5+\chi_{11}|^2
&&& k+2&=12 \cr
Z^{(E_8)} &=|\chi_{1}+\chi_{11}+\chi_{19}+\chi_{29}|^2+
|\chi_{7}+\chi_{13}+\chi_{17}+\chi_{23}|^2 &&&k+2&=30\ . \cr }}
This pattern reflects the existence of an underlying ``extended"
chiral algebra, containing the current algebra $\slh(2)$ as a 
subalgebra. The combinations $\hat\chi_\Ga = 
\sum_{\Gl\in \hT_\Ga} \chi_\Gl$ that appear 
in \Vb\ are %decompositions of 
characters  of the extended algebra decomposed
into irreducible characters of $\slh(2)$.  Let us denote 
$S_{\Gl\Gm}$, resp $\bS_{\Ga\Gb}$, the matrices of modular 
transformations of the two sets of characters
\eqn\Vc{\eqalign{\chi_\Gl(\tilde q)&=\sum_\Gm S_{\Gl\Gm}\chi_\Gm(q)\cr
\hat\chi_\Ga(\tilde q)&=\sum_\Gb \bS_{\Ga\Gb}\hat \chi_\Gb(q)\ ,\cr }}
where $\tilde q=e^{{-2i\pi \over \tau}}$. 
One has $\bS_{\Ga\Gb}=\sum_{\Gl\in \hT_{\Ga}} S_{\Gl\Gm}$, 
$\forall \Gm\in\hT_{\Gb}$. 
The quantum dimensions of the representations are %defined as 
the ratios $\bD_\Ga=\bS_{\Ga 1}/\bS_{11}$ and
$D_\Gl=S_{\Gl 1}/S_{11}$. 

It has been observed in \refs{\PZun,\PZde} that there is a second
manifestation of the $ADE$ diagrams hidden in the structure of the
operator algebra. For the theories \Vb,  
one proves that the fusion coefficients $\bN_{\Ga\Gb}^{\ \ \Gc}$ 
of the extended algebra satisfy
\eqn\Vd{\bN_{\Ga\Gb}^{\ \ \Gc}= \sqrt{\bD_\Ga\over D_\Gl}
\sqrt{\bD_\Gb\over D_\Gm} \sum_{\Gn\in \hT_\Gc} 
M_{\Gl\Gm}^{\ \ \Gn}\sqrt{D_\Gn\over 
\bD_\Gc}\ ,\qquad\qquad \forall \Gl\in\hT_\Ga, \ \Gm\in\hT_\Gb\ , }
 where $M$ are the structure constants of the Pasquier algebra 
of the relevant Dynkin diagram. (For the sake of simplicity, 
we assume here and in the rest of the discussion that 
none of the exponents has a multiplicity  larger than 1: this excludes 
the $D_{{\rm even}}$ case. The cases with multiplicities require 
a more elaborated labelling, see \PZde). 

This equation has several interesting consequences. 
First, since the matrix $\bN$ is diagonalized by the $\bS$ matrix, 
according to the Verlinde formula, it follows from \Vd\ 
that $Y_\Gl:=
 {\bS_{\Ga\Gd}\over \bS_{1\Gd}} \sqrt{D_\Gl\over \bD_\Ga}$,
where $\hT_\Ga$ is the block containing $\Gl$  
 and   $\Gd$ is  any representation of the
extended algebra, %is an eigenvector 
forms a one-dimensional representation of the $M$ algebra, 
i.e. $Y_\Gl \, Y_\Gm=\sum_\Gn M_{\Gl\Gm}^{\ \ \Gn}Y_\Gn$, 
and may thus be identified with some $ {\psi^{(\Gl)}_d\over \psi^{(1)}_d
}$, for some  vertex $d$
\eqn\Ve{{\psi^{(\Gl)}_d\over \psi^{(1)}_d  }
=  {\bS_{\Ga\Gd}\over \bS_{1\Gd}} \sqrt{D_\Gl\over \bD_\Ga} \ .}
%with no summation over $d$ or $\Gd$. 
%
In particular, the Krein parameter of the Pasquier algebra 
reads
\eqn\Vf{\hk_\Gl=D_\Gl \bD_\Ga, \qquad{\rm if }\quad \Gl\in \hT_\Ga}
to be compared with the formula $\hk_\Gl=D_\Gl^2$ of sect. 3.1, 
example 2, valid for the fusion algebras, i.e. for the $A$ cases
for which the blocks $T_\Ga$ contain only one exponent.
%in which the extended algebra 
Let $T$ denote the subset of vertices $d$ for which \Ve\ holds. 
Each of them may be identified with a weight  $\Gd$ of the 
extended algebra.
Further analysis \PZde\  reveals that: \par \noindent
1) $\forall d\in T$, $\Gd$ the 
corresponding extended weight, and for $\Gl \in \hT_\Ga$ one has
\eqn\Vg{{\psi_d^{(\Gl)}\over\psi_1^{(\Gl)}} = {\bS_{\Gd\Ga}\over \bS_{1\Ga}}
\qquad {\rm and}\qquad \psi_1^{(\Gl)}= S_{1\Gl} \bS_{1\Ga}\ ;}
2) one is precisely in the conditions of sect. 3.2: the set $T$ 
defines a C-subalgebra of the $N$ algebra. In the cases of \Vb\
discussed here, the $M$ and $N$ structure constants are 
non negative (see Fact 3 of sect. 1). One may apply the theorem
of Bannai and Ito: the dual subalgebra is associated with a
special set $\hT$ which is the block of the identity representation
and it defines a partition of the set of exponents into classes $\hT_\Ga$. 
Finally equation \Vd\ may be seen to be equivalent to 
equation \IIi\ (or rather its dual), if one takes into account 
the change of normalization between the $q_{\Gl\Gm}^{\ \ \Gn}$
and $M_{\Gl\Gm}^{\ \ \Gn}$ structure constants and the explicit expressions
of the Krein parameters \Vf. 

Thus behind the modular invariants \Vd, there is again a structure of
C-algebras and subalgebras. This had been first pointed out in 
\DFZ, and then the more systematic discussion of \PZde\ has shown that this 
follows from the basic equation \Vd, and that it yields a way 
to {\it determine}  the expressions of some eigenvectors
 from conformal data (quantum dimensions). 

%%%%%%%%%%%%%%%%%%%%%%%%%%%%%%%%%%%%%%%%%%%%%%%%%%%%%%%%%%%%%%%%

\newsec{Conclusion and perspectives}
\noindent 
The purpose of this lecture was to present the concept 
of C-algebra and to illustrate its utility in two 
contexts: the discussion of reflection groups and their 
foldings on the one hand, and the structure of conformal 
field theories, on the other. 

Note that these two 
seemingly disparate problems are in fact related 
in the framework of 2-dimensional topological field theories. For those
theories, or at least for those that are obtained by twisting a 
$\CN=2$ superconformal coset field theory, one has two approaches
at one's disposal: the discussion of the (super)conformal 
field theory following lines analogous to the  discussion of sect. 5; 
and the analysis of the Witten-Dijkgraaf-Verlinde-Verlinde equations
\WDVV, for which Dubrovin \Dub\ has shown the appearance of monodromy
groups generated by reflections. 
In fact the concept of C-algebra seems to be underlying  in a natural
way the whole discussion of topological field theories. 

Note also that in the two discussions of the previous sections, 
the same C-algebras (based on the Pasquier algebra of the 
Dynkin diagrams) have been used in two different ways: 
in one case (folding), we have been looking at the C-subalgebras
of the $M$ algebra (subject to some constraint); in the other
(rcft), it is rather some subalgebra of the $N$ algebra
that has determined the special set $T$ of vertices, 
and by duality the blocks $T_\Ga$. 

One issue that requires clarification is the role of positivity. 
We have from the start restricted our attention to the subcases
of the $ADE$ list that have certain positivity properties (see sect. 2).
The main benefit has been the possibility to use the theorem of Bannai 
and Ito (sect. 3.2). It is possible to relax the positivity 
assumption in the discussion of folding of graphs and groups:
what is really crucial is eq. \IIh,  see \Tani.
In the case of rcft, it is less clear how to proceed and  what replaces
\Vd. In that case, however, we know that any theory with a non block 
diagonal modular invariant (e.g. \Izb{b}) may be obtained from 
a block diagonal one (\Izb{a} in that case) by %a $\IZ_2$ twist 
 an automorphism of the fusion algebra \MSDV. The proper incorporation of that 
fact in the present considerations remains to be done. 

As already mentionned, the very good news is 
that all this discussion is not limited to the $sl(2)$-ADE cases 
to which I have restricted myself here for simplicity. On the contrary, 
both the  folding of generalized   Dynkin diagrams 
associated with $sl(N)$ and the block structure 
 of $\slh(N)$ RCFT may be discussed in quite general terms. 
The C-algebra method enables one to find in a fairly 
systematic way the possible foldings of these generalized diagrams
that respect some general properties, and in the second context, 
it gives non trivial relations between 
conformal data (fusion coefficients and quantum dimensions)  
and eigenvectors of the adjacency matrices. It may even enable
one to {\it construct } the graph from these data. 
See \Tani\ for the former subject and \PZde\ for the latter.

%%%%%%%%%%%%%%%%%%%%%%%%%%%%%%%%%%%%%%%%%%%%%%%%%%%%%%%%%%%%%%%%
\vskip1truecm
\noindent {\bf Acknowledgements} It is a pleasure to thank the
organisers of this symposium for a very interesting and profitable 
meeting. 
I have benefited from an interesting conversation with T. Yano. 
Stimulating discussions with M. Bauer are also gratefully acknowledged. 
Finally I want to recall that most of the results presented here 
have been worked out in collaborations with P. Di Francesco and
V. Petkova. 
 
%%%%%%%%%%%%%%%%%%%%%%%%%%%%%%%%%%%%%%%%%%%%%%%%%%%%%%%%%%%%%%%%
\footatend%\vfill\supereject
\vskip15truemm
\immediate\closeout\rfile\writestoppt
\baselineskip=14pt\centerline{{\bf References}}\bigskip{\frenchspacing%
\parindent=20pt\escapechar=` \input refs.tmp\vfill\eject}\nonfrenchspacing
\bye